# GPU Accelerated Finite Element Assembly with Runtime Compilation

Tao Cui
Academy of Mathematics and
Systems Science, Chinese
Academy of Sciences,
China
tcui@lsec.cc.ac.cn

Xiahu Guo
STFC, Daresbury Laboratory,
UK
xiaohu.guo@stfc.ac.uk

Hui Liu
Department of Chemical and
Petroleum Engineering,
University of Calgary,
Canada
hui.j.liu@ucalgary.ca

## ABSTRACT

In recent years, high performance scientific computing on graphics processing units (GPUs) have gained widespread acceptance. These devices are designed to offer massively parallel threads for running code with general purpose. There are many researches focus on finite element method with GPUs. However, most of the works are specific to certain problems and applications. Some works propose methods for finite element assembly that is general for a wide range of finite element models. But the development of finite element code is dependent on the hardware architectures. It is usually complicated and error prone using the libraries provided by the hardware vendors. In this paper, we present architecture and implementation of finite element assembly for partial differential equations (PDEs) based on symbolic computation and runtime compilation technique on GPU. User friendly programming interface with symbolic computation is provided. At the same time, high computational efficiency is achieved by using runtime compilation technique. As far as we know, it is the first work using this technique to accelerate finite element assembly for solving PDEs. Experiments show that a one to two orders of speedup is achieved for the problems studied in the paper.

## 1 INTRODUCTION

General-purpose computing on graphics processing units (GPGPU) has been developed rapidly in the recent years. Applications in many areas like machine learning, molecular dynamics, computational chemistry, medical imaging and seismic exploration are taking the advantage of GPGPU. Most of the current researches on finite element method with GPU mainly focus on solving the system of linear equations [1-7]. Some of the research works focus on solving specific problems with GPU accelerated finite element methods [8-12]. Several general algorithms for a wide range of finite element models are proposed in [13] and the papers cited in. But the development of finite element code is dependent on the hardware architectures which is usually complicated and error prone using the libraries provided by the hardware vendors. In this paper, we present architecture and implementation of the finite element assembly for partial differential equations (PDE) based on symbolic computation and runtime compilation technique on GPU. This symbolic-numeric architecture is first proposed in [16, 17] for solving PDEs on CPU. Our work adopts the symbolic-numeric paradigm and extends it to GPGPU with runtime compilation. In our system, a user friendly programming interface with symbolic computation is provided. At the same time, high computation efficiency is achieved by using the runtime GPU kernels compilation library NVRTC provided in the CUDA libraries which is first released with CUDA 7 in 2015. CUDA is a parallel computing platform and application programming interface (API) model created by Nvidia. As far as we know, it is the first work using this technique on GPU to accelerate finite element assembly for solving PDEs.

The main contribution of the paper is

1. The mathematical functions and finite element weak forms are expressed in symbols which are close to their mathematical expressions. It is human friendly and very easy to understand.

2. The expressions in finite element method are simplified and optimized by symbolic manipulation before compiling. The final expressions are compiled during runtime, and further optimization can be achieved on target devices (CPU, GPU) using runtime compilation technique.

3. Device independent implementation is achieved with high portability. End users only focus on the mathematical details of the partial differential equations with finite element method. It is not required to the end users to know the compilation details on different device architectures.

The rest of the paper is organized as follows. Section 2 summarizes the related work and the state of the art. Section 3 presents the overview of the design of the system. Section 4 describes the PDE problem and symbolic computation in FEM. Runtime compilation for GPU compute kernels is presented in Section 5. Experiments and results are discussed in Section 6. Finally, Section 7 concludes the paper.

## 2   RELATED WORK

We focus on the weak form construction and finite element assembly step in FEM. Solving the system of linear equations is outside of our consideration in this paper. Most of the related works for finite element assembly are proposed for GPU devices in the recent years, for example parallel threads run for nonzero value in the global linear system [9], elements with graph coloring partition [18, 19], local matrix approach [20] and coordinate list format of the global matrix [21, 22]. The comparison of the above mentioned methods can be found in [13]. The symmetry of local mass and stiffness matrices on a GPU is investigated in [23]. The architecture of the whole finite element steps are discussed in [24]. Most of the assembly algorithms mentioned above can be used as the backend assembly process in the system proposed in this paper since the system is designed to be able to easily provide transparent assembly algorithm and device architecture support for the end users.

Symbolic computation has a long history in scientific computing which refers to the study and development of algorithms and software for manipulating mathematical expressions and other mathematical objects, such as simplification of expressions, differentiation using chain rule, polynomial factorization, indefinite integration, etc. Symbolic computation is first adopted for expressing finite element functions and weak forms in [16] for solving PDEs using Java language with just-in-time compilation on CPU and later extended to other languages like C++ and Python with a cloud platform enabled in [17]. The adoption of symbolic computation for FEM in [16, 17] is motivated by solving PDE based inverse problems with several newly proposed methods in [25-27] which involve significant amount of symbolic manipulations of the mathematical expressions. In order to evaluate the resulting expressions as fast as possible, JIT compilation technique is introduced together with symbolic computation. The combination of symbolic computation and just-in-time compilation in [16, 17] opens up a new symbolic-numeric paradigm for solving PDEs. The symbolic-numeric paradigm is adopted in our work and extended to GPGPU with runtime compilation using NVRTC provide d by CUDA library. Similar works are done in [14, 15, and 28]. In their work, instead of using symbolic computation, a domain specific language (DSL) is invented and runtime low-level language source code generation is used for runtime compilation to machine code. The DSL approach is not able to take the advantage of compile time grammar and type check in the mature low-level languages such as C++, Python and Java. The way of source code generation for runtime compilation also suffers from the compilation overhead in static compiler, for example, spawning separate processes and disk I/O.

As far as we know, this is the first paper for finite element assembly utilizing symbolic computation and NVRTC for runtime compilation.

## 3   DESIGN OVERVIEW

The architecture of the system is shown in Fig. 1. Starting from expressing the finite element weak forms and functions for the coefficients and right hand side (RHS) by symbolic objects, a system of linear equations can be computed by the system shown in Fig. 1 using GPU devices. An alternative way of expressing the weak forms and functions is the use of an online script which is parsed by the symbolic library and converted to symbolic objects eventually.

Symbolic expressions of the functions and weak forms serve as a template and the symbols in the expressions will be substituted by the specific expressions of the shape functions during the finite element assembly process. A library of the common shape functions and finite elements is developed for the end user to reduce the amount of coding work.

The GPU compute kernels are generated base on the kernel templates and the expressions from symbolic objects. The kernels are further compiled to PTX (a low-level parallel thread execution virtual machine defined for CUDA). Then, PTX programs are translated at install time to the target hardware instruction set. Numerical integration in finite element codes is performed in the GPU threads to create entries for element stiffness matrices and element load vectors that are further assembled into the global stiffness matrix and global load vector.

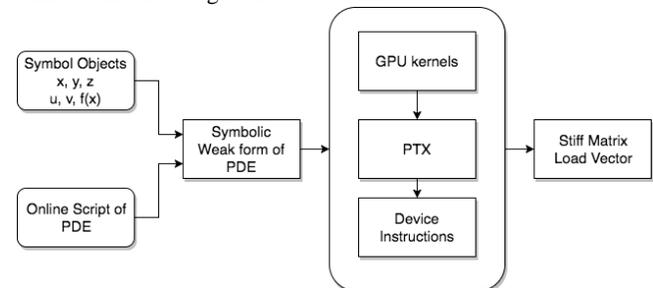

**Fig. 1 Architecture of the symbolic-numeric system on GPU**

## 4   SYMBOLIC COMPUATION IN FEM

### 4.1   Finite Element Method

Without loss of generality, we consider the generalized Helmholtz equation as the example in our paper,

$$-\nabla \cdot \left( \sigma(x,y)\nabla u(x,y) \right) + \lambda u(x,y) = f(x,y),$$
$$(x,y) \in \Omega \tag{1}$$

Where $u(x,y)$ is the solution over a domain $\Omega$. $f(x,y)$ is a given right hand side (RHS). $\sigma(x,y)$ is a symmetric, positive definite matrix and $\lambda$ is a strictly positive constant. Neumann boundary condition on the boundary $\partial\Omega$ of the domain $\Omega$ is considered for simplicity. The weak form of Eq. (1) is,

$$(\nabla v, \sigma \nabla u) + \lambda(v, u) = (v, f), \tag{2}$$

Where $(\cdot, \cdot)$ denotes the inner product over the domain $\Omega$. By choosing the discrete space and piecewise linear finite elements, a system of linear equation is obtained for the weak form,

$$A\vec{x} = \vec{b}, \tag{3}$$

Where the entry $x_i$ of vector $\vec{x}$ denotes the approximation of u on node $v_i$ and i ranges from 1 to the total number of nodes N. Each entry $A_{ij}$ of the matrix A is assembled from all elements that contain both nodes $v_i$ and $v_j$. Similarly each entry $b_j$ of $\vec{b}$ is assembled from all elements that contain $v_i$.

## 4.2  Weak Form Symbolic Representation

Symbolic computation is introduced to express the weak form of a PDE, for example Eq. (2). The coordinates, coefficients, right hand side and shape functions are defined by symbols. The derivatives of the shape functions are obtained by symbolic manipulations. As an example, the following code piece shows an implementation with C++ and GiNaC[30] library.

```
Symbol x("x"), y("y");
ex f = -2*(x*x + y*y) + 36;
matrix sigma = {{1, -x-y},{x+y, 1}};
float lambda = 1.0;
FunctionSpace funSpace(mesh, lst(x,y), "Lagrange", 1);
WeakForm wf(funSpace);
wf.build(
  [&](ex u, ex v) { return dot(grad(v,x,y), sigma*grad(u,x,y)) +
lambda*dot(v,u); },
  [&](ex v) { return f*v; }
);
```

The coordinates (x,y) and coefficient $\sigma(x,y)$ and the RHS $f(x,y)$ are defined as symbolic expressions. The class FunctionSpace wraps the mesh, coordinate and shape functions. The class WeakForm represents Eq. (2) by passing the object of FunctionSpace and the expressions of LHS and RHS of Eq. (2) using lambda functions.

By introducing symbolic computation into FEM, we are able to write the weak forms and related functions for solving PDEs with mathematical appealing application programming interfaces (APIs). The weak forms and functions are further compiled at runtime for numerical computation.

## 5  RUNTIME COMPILATION

Runtime compilation is a way of executing computer code that involves compilation during execution of a program – at run time – rather than prior to execution. NVRTC, originally released with CUDA 7 in 2015 is a Runtime Compilation library which enables compilation of CUDA C++ device code at run time. It provides optimizations and performance not possible in a purely offline static compilation. The overhead associated with spawning separate processes, disk I/O etc. in an offline static compilation is eliminated. Furthermore, the application deployment becomes simple since the related offline compilation tools such as *nvcc* is not required to be installed in the user environment.

Runtime compilation is adopted in our system to compile the generated GPU compute kernels code from symbolic expressions of FEM. There are two ways of generating GPU compute kernels. The first and most common way is based on source code template. A template of the GPU kernels source code that contains template variables which are replaceable at runtime is loaded at the beginning of the execution of a program. By replacing the template variables, the structure of thread blocks, memory hierarchy, loop and array boundaries of the GPU kernels are generated to help the compiler generate better PTX and device instructions. The PTX defines a virtual machine and instruction set architecture (ISA) for general purpose parallel thread execution. PTX programs are translated at install time to the target hardware instruction set. The second way is based on the abstract syntax tree (AST) of the GPU compute kernels for FEM. No source code template is needed. PTX is generated from the AST directly.

Many different finite element assembly algorithms for GPU compute kernels are discussed in Section 2. With the development of hardware devices new compute capabilities are added to the new hardware. For example, atomic operations on shared and global memory are available on devices of compute capability 2.x and higher. Atomic operations are operations which are performed without interference from any other threads. They are often used to prevent race conditions. The atomicAdd() function is an import function for the assembly of finite element local and global matrices. In order to take the advantage of the new atomic operations we propose an algorithm using the atomicAdd() function on both shared and global memory.

At the same time, the number of cores in GPU devices is increased significantly in the newly released devices, for example Tesla V100 has 5120 cores. Our algorithm is also designed to utilize the large number of cores to achieve higher parallel process capability. Fig. 2 shows a 3D thread block for the proposed finite element assembly algorithm. A thread of a block is represented by a small cube in Fig. 2. In the x direction, each thread evaluates the integrand at a numerical integration point. In the y direction, an entry of the local stiffness matrix is computed. In the z direction, a finite element is processed.

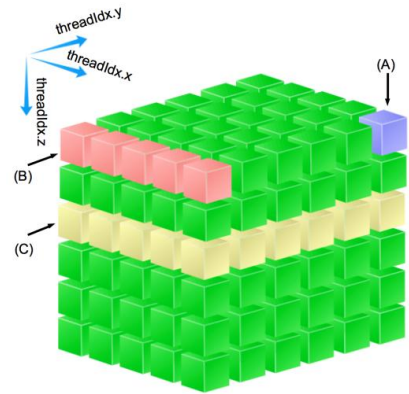

**Fig. 2 A 3D thread block for finite element assembly: (A) A single thread which evaluates the integrand on an single integration point. (B) All the threads in the threadIdx.x direction which compute an entry in the local stiffness matrix. (C) All the threads in the threadIdx.x and threadIdx.y direction which compute the local stiffness matrix for an element. The dimension of threadIdx.z is the number of elements computed in the thread block.**

Fig. 3 shows a memory hierarchy for the GPU compute kernels. The coordinates X, Y (for the example 2D problem) and the global indices of the nodes gIdx are stored in separated arrays in the global memory. The values of the global memories (global arrays) are copied from host memory to device memory during runtime. In each block, the global arrays are copied to the shared memories (shared arrays) corresponding to the coordinates and the global indices of the nodes for the elements processed in the block. The local stiffness matrix and load vector are assembled into shared memory in each block. Finally, the local stiffness matrix and load vector are assembled into global stiffness matrix and load vector.

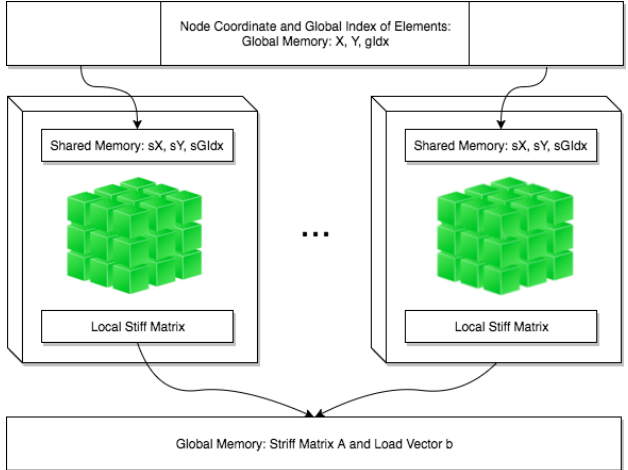

**Fig. 3 Memory hierarchies of the finite element assembly GPU kernels**

The pseudo code of the generated GPU compute kernel for dense stiffness matrix and load vector is listed in Algorithm 1. The details of the algorithm are explained below.

**Algorithm 1. Assembly algorithm using atomicAdd()**

1  INPUT: the array of coordinates X, Y of all elements, the global index gIdx of nodes of all elements.

2  OUTPUT: the global stiffness matrix A and load vector b.

3  Copy global memory X, Y and gIdx to shared memory for the elements in the current thread block.

4  Call _syncthreads().

5  Allocate memory for local stiffness matrix and load vector.

6  Pass coordinates and constants of integration into generated integrand function.

7  Call atomicAdd() to add the value to the local stiffness matrix and load vector.

8  Call _syncthreads().

9  Call atomicAdd() to add the values to the global stiffness matrix and load vector.

Lines 1 and 2 are the input and out values. The values of the coordinate arrays X and Y are computed in this way. For the example 2D problem, there are the three nodes on each element, the coordinates of the three nodes of the first element are stored consecutively in X and Y in the indices 0 to 2. Then the coordinates of the next elements are stored in the indices 3 to 5 and so on. Similarly, gIdx stores the global index of the nodes corresponding to the indices of the nodes in X and Y.

Lines 3 and 4 are used to copy the values from the global memory to shared memory. Threads synchronization is called to ensure all the values are ready for all the threads.

Lines 5-7 compute the local stiffness matrix and load vector. The arrangement of the threads in a block and memory hierarchies are described in Fig. 2 and Fig. 3. The integrand is a separate kernel function generated from a kernel function template by replacing the template variables with the expressions obtained from symbolic expression of the weak form.

Lines 8 and 9 wait for the completion of the local stiffness matrix and load vector. Then, the local values are added to the global stiffness matrix and load vector using the atomicAdd() method operated on global memory.

For sparse stiffness matrix, the differences are the input/output values and the line 9 in algorithm 1. Suppose N is the total number of nodes. Instead of an N by N dense matrix for the output A, the dimension of A is changed to N by MAX_NZ, which is the maximum number of non-zero elements in the rows of matrix A. MAX_NZ is usually obtained from the mesh structure and finite element type. In the 2D example, MAX_NZ is 7 since a node has maximum 6 neighbor nodes and the non-zero elements in a row of A is the node itself plus the 6 neighbors. Two more input arrays are passed into the compute kernel, gNbrNodeLen and gNbrNodeIdx, which are the arrays that contain the maximum number of non-zero entries of each row of A and the indices of the non-zero entries in each row of A. The actual codes corresponding to line 9 is changed to search for the index in A for a global column index and call the atomicAdd() function to accumulate the values into the stiffness matrix A.

## 6  EXPERIMENT

### 6.1  Environment Setup

The hardware used in our experiments is listed in Table 1. The GPU implementation of the system is compared with an Intel® Xeon® EE5-2630 CPU implementation for the finite element assembly using the similar technique, symbolic computation and Just-in-Time (JIT) compilation on CPU with LLVM project [29]. CUDA 8 library is used for Tesla M2090 and Tesla K20. CUDA 9 library is used for Tesla V100.

**Table 1: CPU/GPU Device Specification**

| Device | #Cores | Memory |
|---|---|---|
| Intel® Xeon® EE5-2630 v4 @ 2.20GHz | N/A | 256G |
| Tesla M2090 | 512 | 5G |
| Tesla K20 | 2496 | 5G |
| Tesla V100 | 5120 | 16G |

### 6.2 Experiment Using Dense Stiffness Matrix

Fig. 4 shows the result of the first experiment using dense matrix on CPU (Intel® Xeon® EE5-2630) and GPU (Tesla V100).

The assembly time on GPU is decomposed into the sum of kernel time (Tesla V100 kernel), device memory allocate time (cuMemAlloc), data copy time form host to device memory (Host2Device) and data copy time form device back to host memory (Device2Host). From the result we can see that most of the time on GPU is spent on the data copy from device to host memory. It should be noted that for mesh with size 150*150, it uses even longer time than the time on GPU.

The kernel time for all the grid sizes is actually under 1 millisecond. The mesh size cannot be set to larger tan 150*150 since dense stiffness matrix is used and the dense stiffness matrix will not fit into GPU memory. We conclude that for problems with relatively small number of elements the advantage of parallel computing on GPU cannot speedup finite element assembly much, for example, in an order of one or two. This motivated us to experiment with problems with larger size using sparse stiffness matrix.

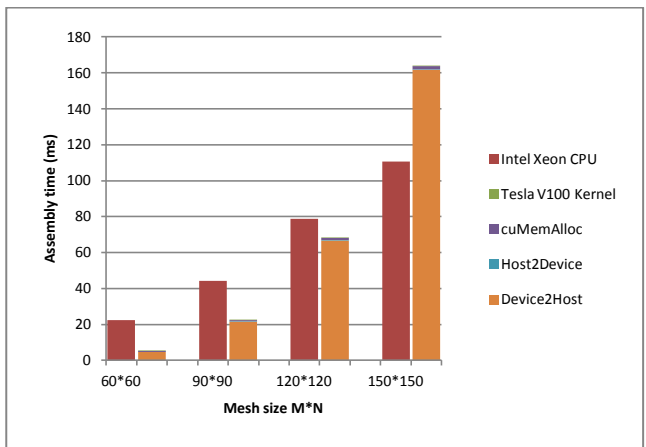

**Fig. 4 Assembly time for dense matrix and vector on CPU and GPU (kernel time and data copy time)**

### 6.3 Experiment Using Sparse Stiffness Matrix

In the second experiment, problems with total number of nodes in the mesh ranges from 10K to 10M are considered. Fig. 5 shows the result of the assembly time on CPU (Intel® Xeon® EE5-2630) and GPUs (Tesla M2090, Tesla K20 and Tesla V100). Here, we mark the horizontal legend as the total number of nodes in the mesh considered instead of mesh size for simplicity. The time for GPU only shows the kernel time since for the problem size considered in this experiment the kernel time takes a considerable portion of the total time consumed for GPU computing. Fig. 6 shows the data copy time from device to host memory on GPUs. The time for other operations is almost negligible.

Fig. 7 shows the speedup for only the time used for kernel computing on GPU compared with the compute time on CPU. Fig. 8 shows the speedup for the total time used on GPU compared with the total time used on CPU. We can see that more than 100x speedup is achieved on Tesla V100 GPU without considering data transfer time. More than 10x speedup on GPU devices is achieved for the case of 10M nodes considering the total time used.

It should be noted that our method is not considering the overhead in traditional static compilation which are typically ranges from hundreds of milliseconds to several seconds which could tear down the speedup significantly.

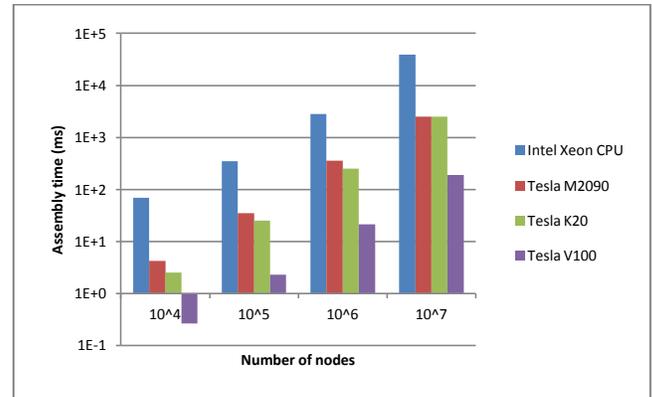

**Fig. 5 Assembly time on different CPU/GPU devices**

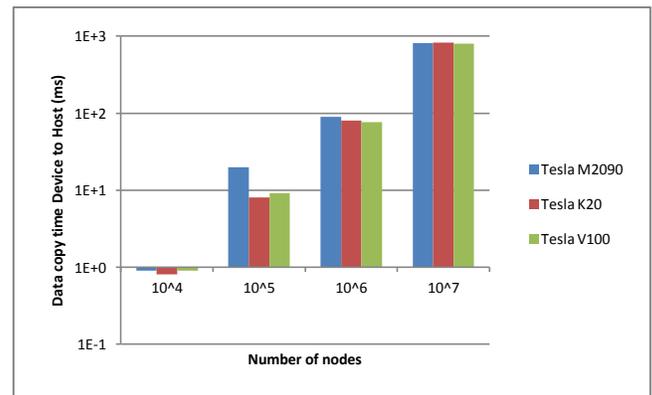

**Fig. 6 Data copy time from device to host memory**

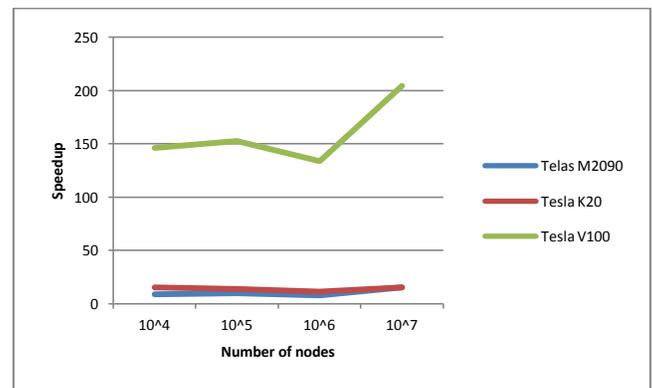

**Fig.7 Speedup for GPU compute kernel vs. CPU**

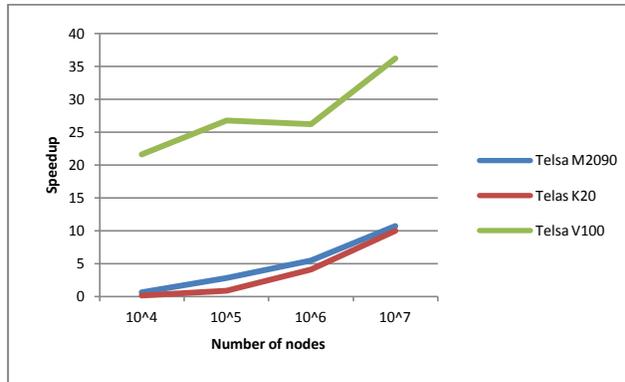

**Fig. 8 Speedup for total time of GPU vs. CPU**

## 7    CONCLUSIONS

Architecture and implementation using symbolic computation and runtime compilation for finite element assembly on GPU are presented in this paper. The proposed system features user friendly API and efficient GPU compute kernels generation for supporting different hardware architecture transparently. A new GPU compute kernel algorithm is presented as the complement of existing algorithms in finite element assembly for the latest Tesla V100 device using atomic add operation on shared and global memory for dense and parse stiffness matrix. Experiment shows an order of one or two speedup is achieved on different GPU devices without considering the overhead in traditional static compilation.

In the future, we are going to consider more complex weak forms and high orders of approximation. In that case, the time for creating the system of linear equations can be higher than the time for its solution since the most time consuming part of the system creation is numerical integration.

## ACKNOWLEDGMENTS

This work was supported by National Key Research and Development Program of China under the grant number 2016YFB0201304, China NSF under the grant numbers 11771440,91530323 and 11101417State Key Laboratory of Scientific and Engineering Computing (LSEC), and National Center for Mathematics and Interdisciplinary Sciences of Chinese Academy of Sciences (NCMIS).